# Theoretical Study of Atomic Structure and Elastic Properties of Branched Silicon Nanowires

Pavel B. Sorokin[*, †, ‡, §] Alexander G. Kvashnin[†], Dmitry G. Kvashnin[†], Pavel V. Avramov[†, §], Alexander S. Fedorov[§], and Leonid A. Chernozatonskii[‡]

[†]Siberian Federal University, 79 Svobodny av., Krasnoyarsk, 660041 Russian Federation

[‡]Emanuel Institute of Biochemical Physics, Russian Academy of Sciences, 4 Kosigina st., Moscow, 119334, Russian Federation

[§]Kirensky Institute of Physics, Russian Academy of Sciences, Akademgorodok, Krasnoyarsk, 660036 Russian Federation

## ABSTRACT

The atomic structure and elastic properties of Y-silicon nanowire junctions of fork- and bough-types were theoretically studied and effective Young modulus were calculated using the Tersoff interatomic potential. In the final stages of bending, new bonds between different parts of the Y-shaped wires are formed. It was found that the stiffness of the nanowires considered can be compared with the stiffness of carbon nanotube Y-junctions.

---

[*] Corresponding author. E-mail: PSorokin@iph.krasn.ru

## Introduction

Nanostructures, such as nanocrystals and nanowires (NW), represent the key building blocks for nanoscale science and technology. Nanowires are the most promising elements of nanotechnology. They can be used as field-effect transistors (FETs) [1, 2], logic gates [3], sensors [4] and more. The effective sizes and electronic properties of the species can be controlled during synthesis in a predictable manner. [5] An ability to obtain perfect silicon nanowires with millimeter length [6] opens new perspectives of using NWs as elements of micromechanical devices. Another perspective technological field is to use branched and hyperbranched nanowires. Conceptually, such structures offer another approach for increasing structural complexity and enable greater functionality. For example, branched NWs might serve as building blocks to design 3D interconnected computing structure. [7]

Dendrite-like nanowires were obtained by Fonseca *et al* [8] by the electron-beam-induced approach. Such structures can be treated as the "fork" Y type branched nanowires (Si-Y-NWs, Fig. 1a) with 90º angle between equal branches. Wang *et al.* [9] proposed a synthetic approach to branched nanowires of various chemical compositions. The single silicon nanowires serve as substrates to deposit gold nanoclusters. After that, a routine technique was used to grow nanowire branches in the same way as the initial wire stem. This method can be used for the obtaining the "bough" Y type branched nanowires (Si-YB-NWs, Fig. 1b) with a 60º angle between the branches and stem and effective wire diameters from 22 to 30 nm. Alivisatos *et al.* have reported a wet synthesis of tetrapod or branched nanocrystals of cadmium telluride and cadmium selenide, controlling the diameter of identical arms. [10]

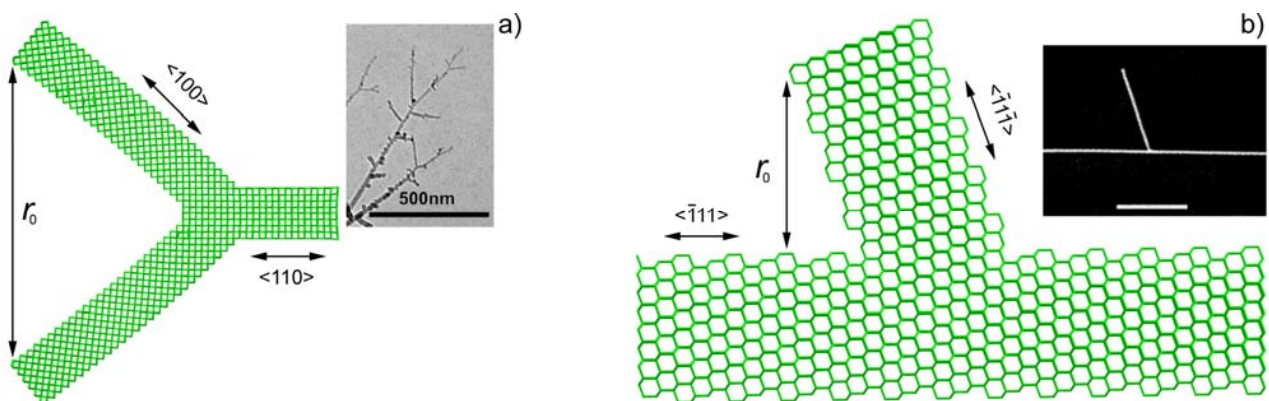

**Fig. 1. The models of Si-Y-NW's: a) "fork"- and b) "bough"- types. Insets: Scanning electron microscopy images of a branched silicon nanowires; a) adapted and reprinted with permission from Ref. 8. Copyright 2005, American Institute of Physics. The scale bar is equal to 500 nm; b) adapted with permission from Ref. 9. Copyright 2004 American Chemical Society. The scale br is equal to 1 μm.**

Though big experimentally activity, only very limited theoretical studies of the branched were carried out. In the paper of Menon *et al.* [11] nanotrees with a stem and perpendicular branches of hypothetical clathrate-like nanowires were investigated and their electronic properties were calculated.

The electronic properties of Si- nanoflowers were studied by Avramov *et al.* [12] The presence of only two theoretical papers devoted this topic can be explained by the fact that the smallest realistic models of the branched NWs contain ~$10^3$ or more atoms, therefore a theoretical study of the proposed objects using ab initio techniques is a great challenge to quantum chemistry.

Presented here is a theoretical study of atomic structure and mechanical properties of the branched silicon nanowires of fork and bough types using the model Tersoff interatomic potential. Using the extended cluster models, the effective Young elastic modulus of the junctions was calculated. It was shown that in contrast to similar Y-junctions of carbon nanotubes, the junctions behave as springs up to contact of the branches. The most drastic changes (including formation of new chemical bonds) in the atomic structure of the branched nanowires caused by the external stress are located in the junction region. The unique mechanical properties of the clusters enable their use as structural units of mechanical nanodevices.

## *Structural models and methods of calculations*

Two possible types of three-terminal branched nanowires of finite lengths were studied. The first type (Y- "fork" configuration, Fig. 1a) has two equivalent branches attached to a stem. This cluster was used to study the dendrite-like nanowires previously obtained experimentally. [8] Unfortunately, no data concerning the atomic structure of the experimental wires are available. To describe the silicon dendrite structures we chose the $\langle 110 \rangle$ orientation of the NW stem and the $\langle 100 \rangle$ orientation of the branches as the most natural structure of the Si-Y-NWs. We assumed that the nanowires form a perfect Si-crystal lattice [13] with unpassivated and hydrogen passivated surfaces.

The "bough" three-terminal Si-YB-NW branched nanowire is formed by connection of a nanowire branch with the stem (Fig. 1b). We used this type of clusters to model the structure of branched silicon nanowires obtained experimentally in the work of Wang *et al.* [9] (see Fig. 1b-inset). It was shown that the stem is oriented along the $\langle \bar{1}11 \rangle$ direction whereas the branch is oriented along the $\langle \bar{1}1\bar{1} \rangle$ direction and was considered as a low-dimensional silicon monocrystal. To study the mechanical properties we designed a number of finite branched Y-shaped "bough"-type nanowires with effective diameters of the stem and branches in the range of 1-2 nm. The Young modulus of the wires with longer branches were calculated using extrapolation of the data obtained for finite systems.

The simulations were performed using molecular mechanics with Tersoff model potential. [14,15] The model potential successfully describes the atomic structure of bulk silicon and silicon nanowires, [16] silicon surfaces, [17] point defects, [18] and their elastic properties. [19,20] Depending on the wire configuration, up to 10000 atoms were involved in the simulations using Broyden-Fletcher-Goldfarb-Shanno (BFGS) [21] optimization technique.

To simulate external pressure, we used the method of constructing an atomic plane [22, 23] with a frozen atomic plane fragment ($S_{fragment}$ = 200 Å$^2$) driven towards a branch in small steps. We used the same plane fragment for the all studied structures for the correct comparison of obtained elastic data. We performed and optimization of the atomic structure of the entire wire at each step. The wire was fixed in transversal direction to avoid shaking the structure. Also, the end of the stem of the "fork" NW and both ends of the stem of "bough" NW were fixed. To calculate the NW's strain energy we chose the pure repulsive potential between the plane and nanowire to avoid nonrealistic bonding between them. The bending strain was defined as $\varepsilon = \frac{\Delta r}{r_0}$ where $\Delta r = r - r_0$, $r_0$ is the distance in unstrained structure between branch ends in the case of "fork" NW and between the end of the branch and the stem in the case of "bough" NW, $r$ is the current distance (Fig. 1).

## *Results and discussion*

The changes in the total energy with respect to the total energy of the initial strain-free configuration reflect the strain energy $E$ as a function of strain $\varepsilon$. The strain energy $E(\varepsilon)$ for the Y-fork NW ($d$ = 17 Å, $L$ = 40 Å, Fig. 2a) displays a quadratic behavior until ε = 0.75. Increasing pressure leads to increased bond lengths between Si atoms at the outer wire surface region and shortened ones in the inner surface region with visible deflection the bond angles from their natural tetrahedral value of 109.471$^o$ (compare Fig. 2b and Fig. 2c). The regions of the main distortion of the crystalline structure are the branch crossing interface (for "fork" Si-Y-NW, bonds between atoms here are shorter and marked by red, see Fig. 2c) and crossing of the branch and stem for Si-NW of bough type. Further increasing of the tension leads to reforming of chemical bonds in the region of branch crossing (ε ≥ ε$_{critical}$, Fig. 2d) and formation of new bonds between the branch ends (ε ≥ ε'$_{critical}$, Fig. 2e). The lattice structure of the other nanowire regions remains practically undistorted.

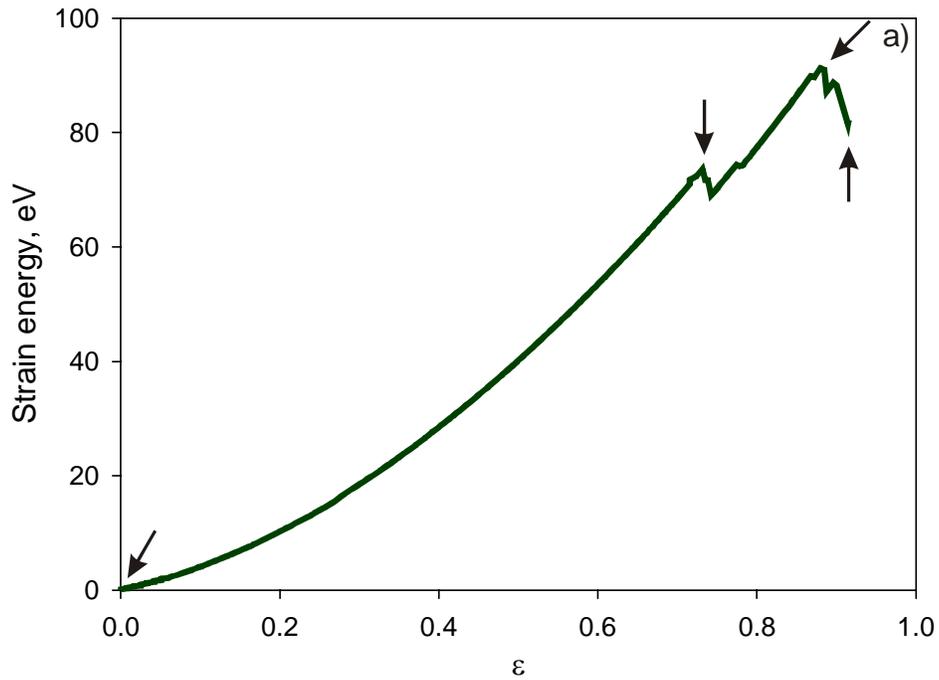

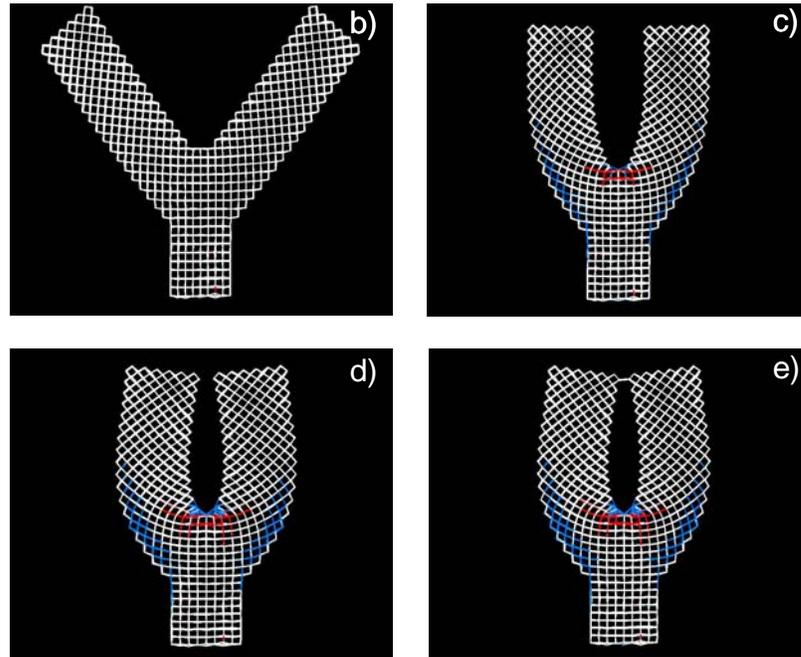

**Fig. 2. Behavior of a Si-NW "fork" junction under loading a) strain energy of the Si-Y-NW ($d = 17.3$ Å, $L = 49.6$ Å) as a function of deformation $\varepsilon$. The arrows correspond to the stages of wire compression: b) unstrained configuration ($\varepsilon = 0$) c) $\varepsilon_{critical} = 0.75$ d) $\varepsilon'_{critical} = 0.87$ before creation of the bonds between the branches and e) $\varepsilon'_{critical} = 0.87$ after creation of the bonds between the branches of the Y-NW. The bonds shown in red have lengths less than 2.26 Å, bonds shown in blue have lengths more than 2.41 Å. (the bond length in the bulk silicon is equal to 2.3516 Å [13]).**

The unloading of the wire in the region $\varepsilon_{critical} > \varepsilon > \varepsilon'_{critical}$ leads to unbending of the wire to a new structure with reformed bonds in the region of branch crossing with slightly higher energy than initial structure ($\Delta E \approx 3$ eV) and larger effective Young modulus (1.9 GPa) than initial structure (1.8 GPa). This effect can be used for hardening Y-wires. For Y-shaped "fork" nanowires, the elastic data ($\varepsilon_{critical}$, $\varepsilon'_{critical}$, $E''$, $Y_{eff}$, $E''/r_0$) is presented in Table 1

**Table 1. Elastic properties of studied Si-Y-NW. d and L are the wire's branch diameter and length respectively, $\varepsilon_{critical}$ is critical strain for reformation of the chemical bonds in the region of branch crossing, $\varepsilon'_{critical}$ is critical strain to form new chemical bonds between the branch ends**

| $d$, Å | $L$, Å | $\varepsilon_{critical}$ | $\varepsilon'_{critical}$ | $E''$, eV | $Y_{eff}$, GPa | $E''/r_0$, eV/Å |
|---|---|---|---|---|---|---|
| | 24.7 | 0.67 | 0.68 | 56.23 | 1.02 | 1.71 |
| | 36.1 | 0.88 | 0.89 | 42.82 | 0.55 | 0.92 |
| **13.4** | 47.3 | 0.83 | 0.84 | 34.41 | 0.33 | 0.55 |
| | 86.8 | 0.94 | 0.98 | 17.19 | 0.11 | 0.19 |
| | 110.1 | 0.74 | 0.96 | 17.63 | 0.08 | 0.14 |
| | 27.5 | 0.43 | 0.93 | 327.40 | 5.98 | 9.98 |
| **17.3** | 49.6 | 0.73 | 0.92 | 179.21 | 1.84 | 3.07 |
| | 68.6 | 0.95 | 0.97 | 102.05 | 0.84 | 1.39 |
| | 76.6 | 0.94 | 0.97 | 112.85 | 0.77 | 1.29 |

We estimated the effective Young modulus of the Si-Y-NW junctions with two different diameters (17.3 Å and 13.4 Å) as $Y_{eff} = \dfrac{E''}{r_0 S}$, where strain energy $E$ was approximated as $E = \dfrac{1}{2} E'' \varepsilon^2$ on the assumption that the loading was mainly created by the atoms of the atomic plane and $S$ is the square of the plane ($S = S_{fragment}$). The increasing of the length of the branches from 24.7 to 110.1 Å and from 27.5 to 76.6 Å leads to decrease in the Young modulus from 1.02 GPa to 0.08 GPa for wires with diameter 13.4 Å and from 5.98 GPa to 0.77 GPa for the 17.3 Å nanowires (Fig. 3). Due to decreasing the relative number of distorted bonds and bond angles with increase of the branch lengths, the effective Young modulus of the wires with longer branches cannot be larger ones for studied systems.

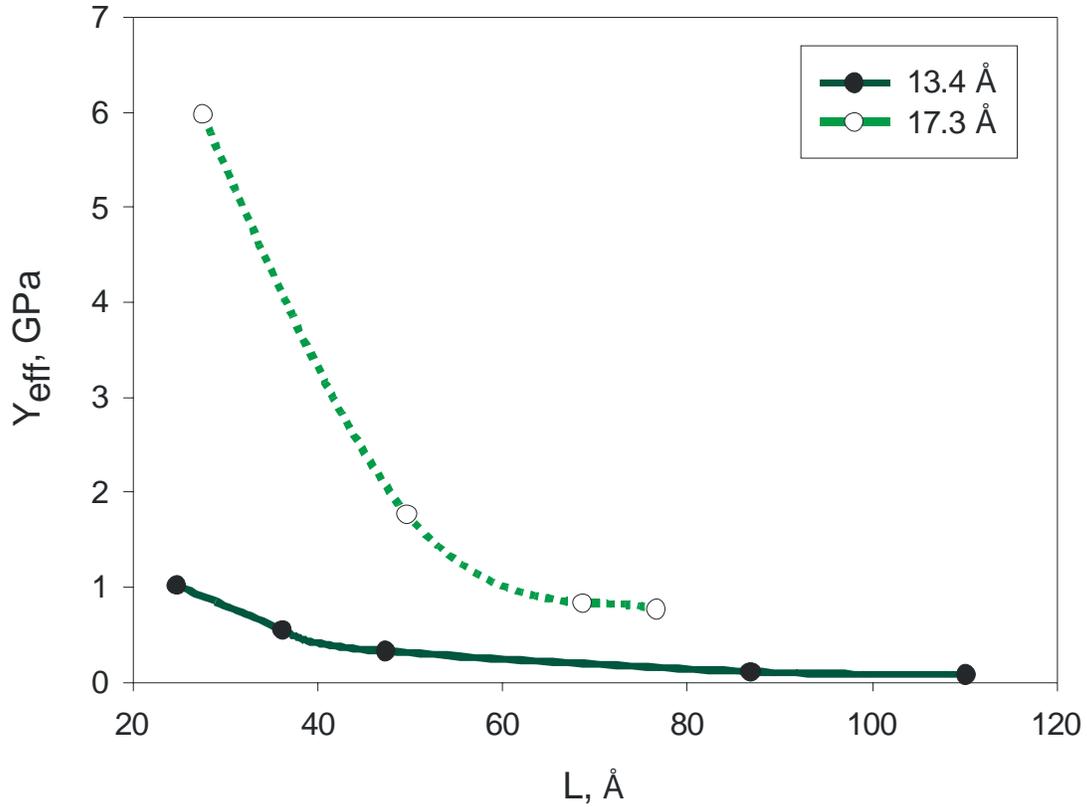

**Fig. 3.** The effective Young's modulus ($Y_{eff}$) for Si-Y-NW as a function of branch length (*L*) of the nanowires. Two curves correspond to different nanowire diameters (13.4 and 17.3 Å).

The unpassivated surface of the wire can influence the mechanical behavior especially at the critical bending values. We studied the Si-Y-NW (*L* = 49.6 Å, *d* = 17.3 Å) with the hydrogen passivated surface. We studied the bending process of the wire and found that due to the low chemical activity of hydrogen atoms, the branches of the wire do not bond with each other at the final deformation stage ($\varepsilon > \varepsilon_{critical}$, Fig. 4a) and, therefore the strain energy of the passivated wire does not fall down and the energy moves up (Fig. 4b). Also the presence of hydrogen atoms increases the Si-Y-NW effective Young modulus (passivated Si-Y-NW: $Y_{eff}$ = 2.58 GPa, unpassivated Si-Y-NW: 1.77 GPa) due to the interaction of neighboring hydrogen atoms during the bending.

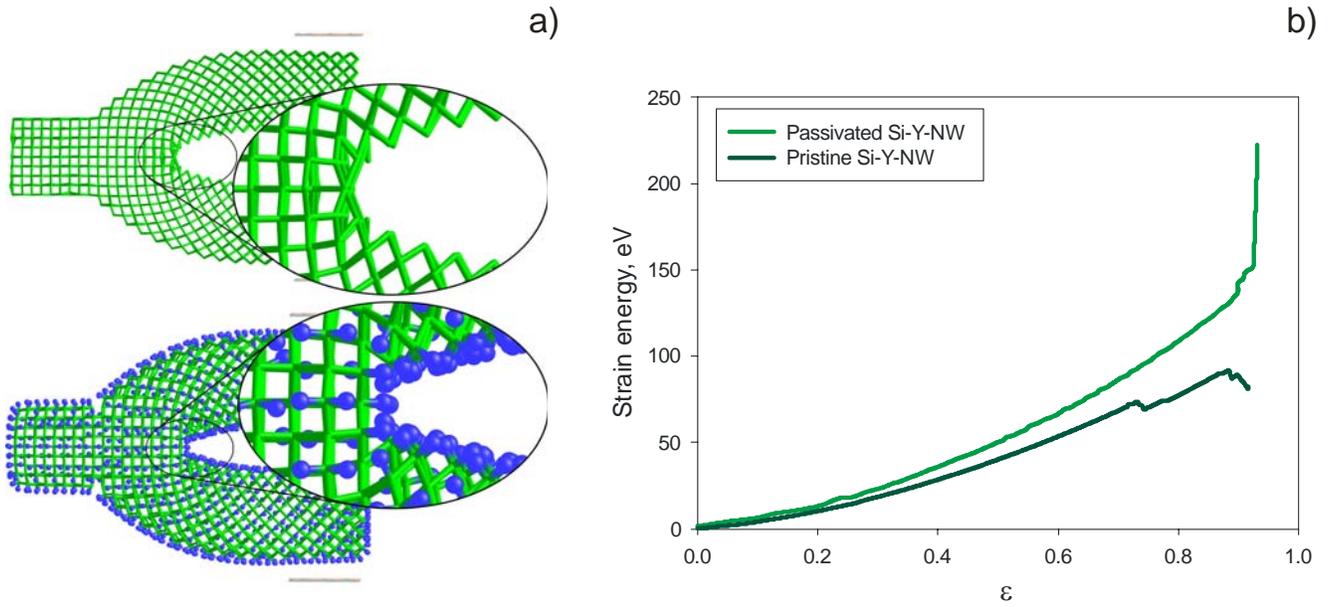

**Fig. 4. The comparison of passivated and unpassivated Si-Y-NW at ε = 0.78: The unpassivated Si-Y-NW bent inelastically due to the reforming of the bonds in the region of branch crossing in the region 0.75 < ε < 0.87, whereas the passivated Si-Y-NW bended elastically.**

To study the effective Young modulus of the Y-shaped "bough"-junctions type nanowires (Si-YB-NWs, Fig. 1b) we calculated nanowires with diameters 9.5 Å, 13.5 Å and 18.8 Å (Fig. 5b-e). To load the stress, the atomic plane was driven toward the single wire branch approaching the stem. Both ends of the stem were fixed. In Fig. 4 the results of applying of the external stress to the typical Si-YB-NW with $d$ = 13.5 Å and $L$ = 57.9 Å are presented. The mechanism of deformation of the Si-YB-NW junctions is more complex than the Si-"fork" one. At the first stages of bending, the deformations of both branch and stem occur. Along with the bending of the branch the stem buckles. This leads to oscillation of the strain energy (Fig. 5a) in the region of $0 < \varepsilon < 0.44$. Pure bending of the branch begins from $\varepsilon = 0.44$ (Fig. 5c) up to $\varepsilon_{critical} = 0.87$ (Fig. 5d and Fig. 4a, light green part of the curve). Further increasing the tension leads to reformation of the bonds in the region of branch and stem junction and formation of new covalent bonds between branch end and the stem (Fig. 5e) with a decrease in the strain energy (Fig. 5(a,e)).

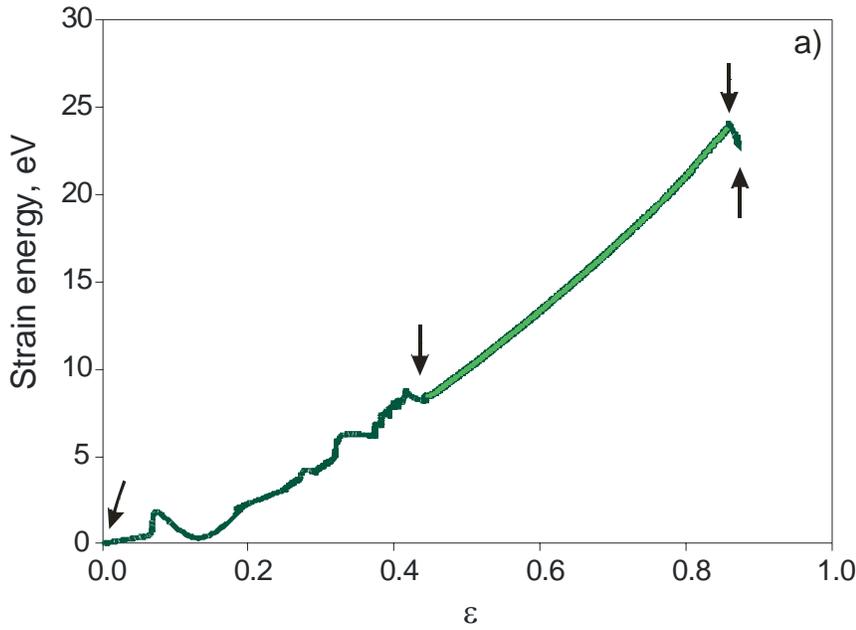

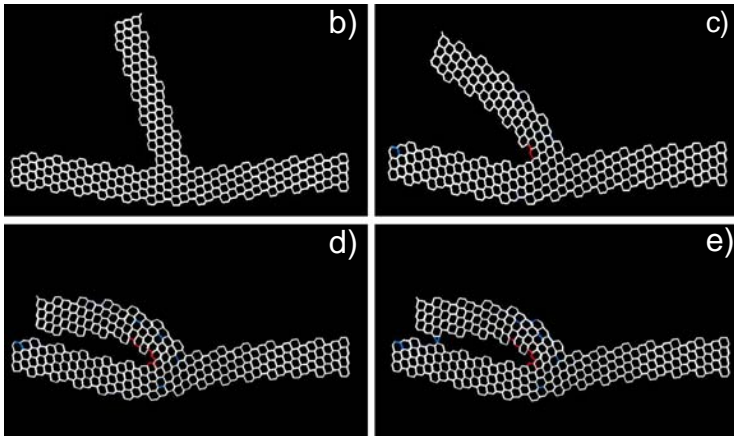

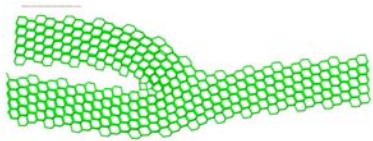

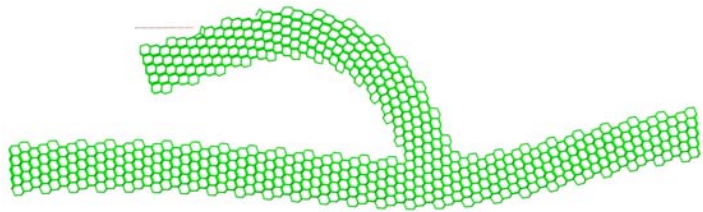

Fig. 5. Behavior of a Si-NW "bough" junction under loading a) strain energy of the Si-YB-NW ($d = 13.5$ Å, $L = 57.9$ Å) as a function of deformation $\varepsilon$. The arrows correspond to the stages of wire compression: b) unstrained configuration ($\varepsilon = 0$) c) $\varepsilon = 0.44$ d) $\varepsilon_{critical} = 0.87$ before the bonding and e) $\varepsilon_{critical} = 0.87$ – after the bonding of the branch and stem. The bonds shown in red have the length less then 2.26 Å, the bonds shown in blue are larger 2.41 Å (the Si-Si bond length in bulk silicon is equal to 2.3516 Å). The comparison between bending behavior of nanowires ($\varepsilon = 0.85$, d = 18.8 Å) with f) short ($L = 60.7$ Å) and g) long ($L = 120.7$ Å) branch lengths.

In the case of Si-YB-NWs with longer branch lengths, a deformation of only the branch part, located far away from the branch-stem crossing, occurs (Figs. 4g and 4f). This effect is caused by irregular stiffness of the Si-YB-NWs.

The results of the calculation of effective Young modulus of Si-YB-NWs are presented in Fig. 6. The elongation of the branches from 45.1 to 107.2 Å, from 38.6 to 117.4 Å and from 44.7 to 120.7 Å leads to a decrease of the Young modulus from 0.49 GPa to 0.22 GPa for the wires with diameter 9.5 Å, from 2.45 GPa to 0.14 GPa for the 13.5 Å wires and from 4.26 GPa to 0.45 GPa for the 18.8 Å nanowires. Due to decreasing distortions in the bond lengths and bond angles with increasing length of the branches, the effective Young modulus of the wires with longer branches cannot be larger the Young modulus of the finite length nanowires. The Y-shaped "bough"-junctions nanowires elastic data is presented in Table 2.

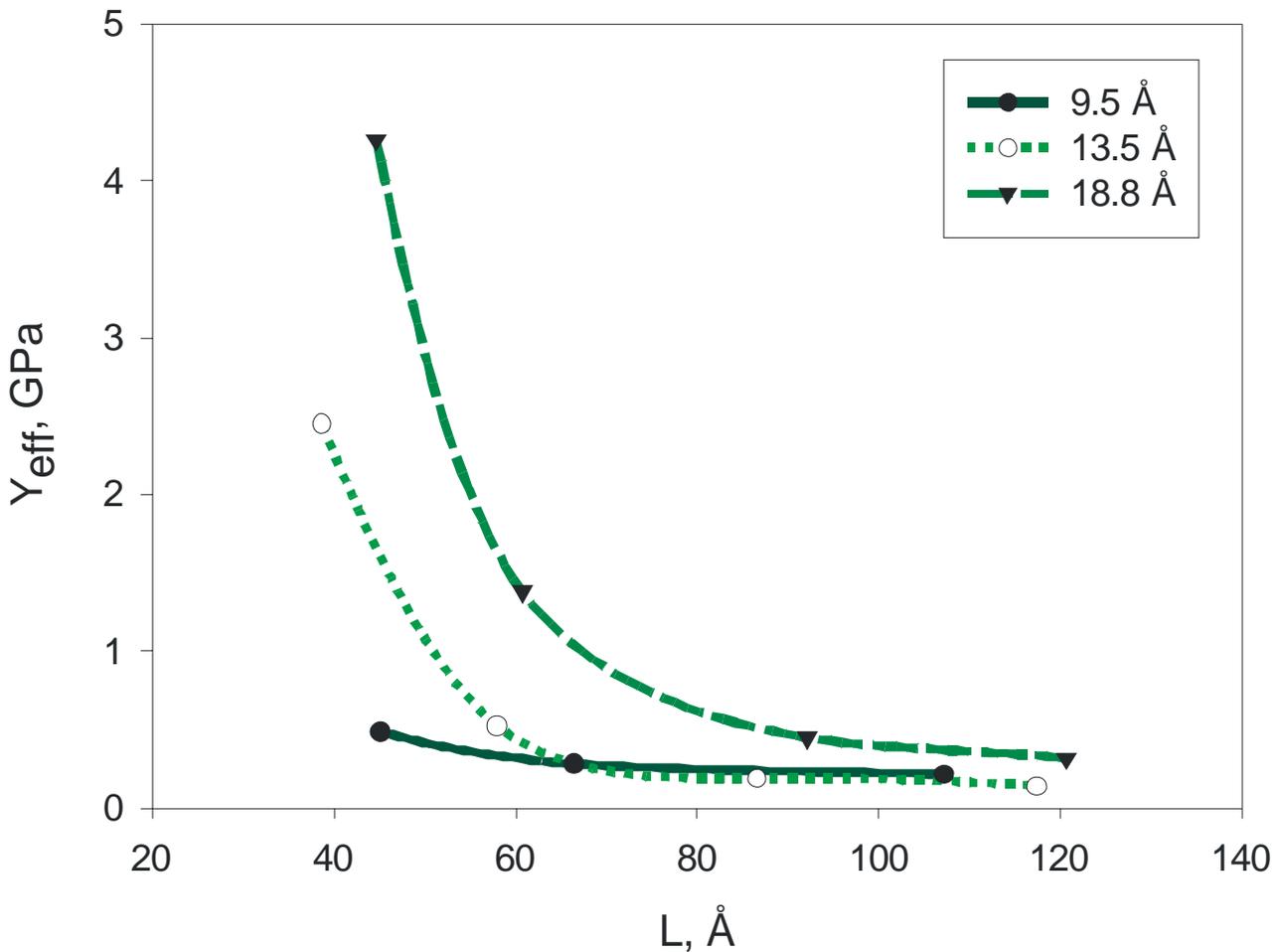

**Fig. 6. The effective Young's modulus ($Y_{eff}$) for Si-YB-NWs as a function of branch length (*L*) of the nanowires. Three curves correspond to different nanowires diamzeters (9.5, 13.5 and 18.8 Å).**

**Table 2. Elastic properties of studied Si-YB-NW. *d* and *L* are the wire's branch diameter and length respectively, $\varepsilon_{critical}$ is critical strain to form new chemical bonds between the branch and the stem**

| *d*, Å | *L*, Å | $\varepsilon_{critical}$ | *E''*, eV | $Y_{eff}$, GPa | $E''/r_0$, eV/Å² |
|---|---|---|---|---|---|
| **9.5** | 45.1 | 0.77 | 30.55 | 0.49 | 0.81 |
|  | 66.4 | 0.84 | 30.32 | 0.28 | 0.47 |
|  | 107.2 | 0.77 | 36.23 | 0.22 | 0.36 |
| **13.5** | 38.6 | 0.66 | 136.45 | 2.45 | 4.08 |
|  | 57.9 | 0.86 | 43.49 | 0.53 | 0.88 |
|  | 86.6 | 0.69 | 26.74 | 0.19 | 0.32 |
|  | 117.4 | 0.78 | 25.54 | 0.14 | 0.24 |
| **18.8** | 44.7 | 0.75 | 245.53 | 4.26 | 7.10 |
|  | 60.7 | 0.90 | 126.90 | 1.38 | 2.30 |
|  | 92.2 | 0.85 | 63.31 | 0.45 | 0.75 |
|  | 120.7 | 0.75 | 59.66 | 0.32 | 0.53 |

In the work of Ref. 23, the effective Young's modulus of the "fork"-type Y junction nanotubes was calculated. For the Y-type single wall nanotubes (*L* = 45 Å and 90 Å, branch diameter *d* = 12.5 Å) and Si-YB-NW (*L* = 45.1 Å and *L* = 107.2 Å, *d* = 9.5 Å, see the Fig. 6 and Table 1) the $E''/r_0$ values are equal to 0.24 eV/Å (Y-SWNT, *L* = 45 Å), 0.12 eV/Å (Y-SWNT, *L* = 90 Å), 0.81 eV/Å (Si-YB-NW, L = 45.1 Å) and 0.36 eV/Å (Si-YB-NW, L = 107.2 Å), respectively. The $E''/r_0$ of double wall carbon nanotube Y junction is slightly larger (0.16 eV/Å, for *L* = 90 Å), than ones for Y-SWNTs, but is less than $E''/r_0$ of the corresponding wire. The nanotube junction elastic data are presented in Table 3 and in the original paper. [23]

**Table 3. Elastic properties of single wall (SWNT) and double wall (DWNT) carbon nanotube "bough" Y junction (from Ref. 22)**

|      | (n,m)-(l,k) | $d$, Å | $L$, Å | $E''$, eV | $Y_{eff}$, GPa | $E''/r_0$, eV/Å |
|------|-------------|--------|--------|-----------|----------------|-----------------|
| **SWNT** | (10,0)-(4,4) | 5.4 | 45.0 | 14 | 1.09 | 0.44 |
|      | (19,0)-(9,9) | 12.2 | 45.0 | 38 | 0.60 | 0.24 |
|      | (19,0)-(9,9) | 12.2 | 90.0 | 34 | 0.30 | 0.12 |
| **DWNT** | (10,0)-(4,4)@(19,0)-(9,9) | 12.2 | 90.0 | 46 | 0.40 | 0.16 |

The unloading of elastically-bent wires should lead to oscillations of branches with a frequency depending on branch lengths and diameters. Therefore such structures can be used as tuning forks with ultrahigh frequencies. We should mention that studied nanowires can be elastically bent under strain up to $\varepsilon = 0.9$ similar to the carbon nanotubes "bough" Y junction investigated earlier [23] and can be used along with them as nanosprings.

The elastic properties of branched Y- silicon nanowires of fork- and bough-types were studied by means of molecular mechanics simulations using the Tersoff model potential, and the effective Young modulus were calculated. During bending, in the inelastic regime the formation of new bonds between different parts of the nanowires in the region of branch crossing was observed. The Si-YB-NWs demonstrate complex mechanisms of deformation under external stress due to the NW's stem buckling in the first steps of bending. Due to irregularity of the branch stiffness, the Young modulus and bending mechanisms of the Si-Y-NWs and Si-YB-NWs depend on the branch lengths. The stiffness of the studied nanowires and nanotubes were compared with the literature. The Si-Y-junctions behave as springs up to contact of branches in contrast to similar Y-junctions of carbon nanotubes. It was found that the stiffness of the wires is much larger due to crystalline structure of the wires. The effective Young modulus of nanowires with longer branch lengths was estimated.

*Acknowledgments*

One of the authors acknowledges encouragement and support of Dr. Keiji Morokuma of Funku Institute for Fundamental Chemistry, Kyoto University. This work was in part supported by a CREST (Core Research for Evolutional Science and Technology) grant in the Area of High Performance Computing for Multi-scale and Multi-physics Phenomena from the Japan Science and Technology Agency (JST) and by Russian Fund for Basic Research (grant n. 09-02-92107 and n. 08-02-01096). All calculations have been performed on the Joint Supercomputer Center of the Russian Academy of Sciences. The geometry of all presented structures was visualized by ChemCraft software.

## *References*

1. Y. Cui, C.M. Lieber, Science, **291**, 851 (2001).

2. Y. Cui, Z.H. Zhong, D.L. Wang, W.U. Wang, and C.M. Lieber, Nano Lett. **3**, 149 (2003).

3. Y, Huang, X. Duan, Y. Cui, L.J. Lauhon, K. Kim, C.M. Lieber, Science **294**, 1313 (2001).

4. Y. Cui, Q. Wei, H. Park, C.M. Lieber, Science **293**, 1289 (2001).

5. A.M. Morales, and C.M. Lieber, Science **279**, 208 (1998).

6. W.I. Park, G. Zheng, X. Jiang, B. Tian, and C.M. Lieber, Nano Letters **8**, 3004 (2008)

7. D. Wang, C.M. Lieber, Nature Mater. **2**, 355 (2003)

8. L.F. Fonseca, O. Resto, and F. Solá, Appl. Phys. Lett. **87**, 113111 (2005)

9. D. Wang, F. Qian, C. Yang, Z. Zhong, C.M. Lieber, Nano Letters **4**, 871 (2004)

10. L. Manna, D.J. Milliron, A. Meisel, E.C. Scher, A.P. Alivisatos, Nano Letters. **5**, 2164 (2005).

11. M. Menon, E. Richter, I. Lee, and P. Raghavan, J. Comp. and Theor. Nanoscience **4**, 252 (2007)

12. P.V. Avramov, L.A. Chernozatonskii, P.B. Sorokin, M.S. Gordon. Nano Lett. **7**, 2063 (2007)

13. R.W.G. Wyckoff, Crystal Structures (Interscience Publishers, Inc., New York, 1948), Vol. 1, in the second edition (John Wiley & Sons, Inc., New York, 1963).

14. J. Tersoff, Phys. Rev. B **38**, 9902 (1988)

15. F. de Brito Mota, J.F. Justo and A. Fazzio, J. Appl. Phys., **86**, 1843 (1999)

16. Y. Zhao, B.I. Yakobson, Phys. Rev. Lett. **91**, 035501 (2003)

17. W. Liu, K. Zhang, H. Xiao, L. Meng, J. Li, G.M. Stocks, J. Zhong, Nanotechnology **18**, 215703 (2007)

18. M. Posselt, F. Gao and H. Bracht, Phys. Rev. B **78**, 035208 (2008)

19. R. Zhu, E. Pan, P.W. Chung, X. Cai, K.M. Liew, A. Buldum, Semicond. Sci. Technol. **21**, 906 (2006)

20. T.Y. Kim, S.S. Han and H.M. Lee, Materials transactions **45**, 1442 (2004)

21. W.H. Press, S.A. Teukolsky, W.T Vetterling B.P. Flannery, Numerical Recipes, 2nd edn. (Cambridge University Press, Cambridge, 1992)

22. L.A. Chernozatonskii and I.V. Ponomareva, JETP Lett. **78**, 327 (2003)

23. E. Belova, L.A. Chernozatonskii, Phys. Rev. B **75**, 073412 (2007)